\documentclass[preprint2]{aastex63}

\usepackage{float}
\usepackage{gensymb}
\usepackage{amsmath}
\usepackage{dblfloatfix}
\usepackage{savesym}
\savesymbol{tablenum}
\usepackage{dirtytalk}
\usepackage{siunitx}
\restoresymbol{SIX}{tablenum}
\usepackage{graphicx}
\shorttitle{Modeling Daily Coronal Spectral Irradiance}
\shortauthors{Schwab et al.}
\graphicspath{{./}{figures/}}

\begin{document}

\title{Modeling the Daily Variations of the Coronal X-ray Spectral Irradiance with Two Temperatures and Two Emission Measures}

\author[0000-0002-1426-6913]{Bennet D. Schwab}
\affiliation{Ann \& H.J. Smead Department of Aerospace Engineering Sciences, University of Colorado at Boulder, 3775 Discovery Dr., Boulder, CO 80303}

\author[0000-0002-2308-6797]{Thomas N. Woods}
\affiliation{Laboratory for Atmospheric and Space Physics, University of Colorado at Boulder, 3665 Discovery Dr., Boulder, CO 80303}

\author[0000-0002-3783-5509]{James Paul Mason}
\affiliation{Laboratory for Atmospheric and Space Physics, University of Colorado at Boulder, 3665 Discovery Dr., Boulder, CO 80303}

\begin{abstract}

The Miniature X-ray Solar Spectrometer (MinXSS-1) CubeSat observed solar X-rays between 0.5 and 10~keV. A two-temperature, two-emission measure model is fit to each daily averaged spectrum. These daily average temperatures and emission measures are plotted against the corresponding daily solar 10.7~cm radio flux (F10.7) value and a linear correlation is found between each that we call the Schwab Woods Mason (SWM) model. The linear trends show that one can estimate the solar spectrum between 0.5~keV and 10~keV based on the F10.7 measurement alone. The cooler temperature component of this model represents the quiescent sun contribution to the spectra and is essentially independent of solar activity, meaning the daily average quiescent sun is accurately described by a single temperature (1.70~MK) regardless of solar intensity and only the emission measure corresponding to this temperature needs to be adjusted for higher or lower solar intensity. The warmer temperature component is shown to represent active region contributions to the spectra and varies between 5~MK to 6~MK. GOES XRS-B data between 1-8 Angstroms is used to validate this model and it is found that the ratio between the SWM model irradiance and the GOES XRS-B irradiance is close to unity on average. MinXSS-1 spectra during quiescent solar conditions have very low counts beyond around 3~keV. The SWM model can generate MinXSS-1 or DAXSS spectra at very high spectral resolution and with extended energy ranges to fill in gaps between measurements and extend predictions back to 1947.
\end{abstract}



\section{Introduction} \label{sec:intro}

There are daily observations of the solar spectral irradiance to study solar physics (e.g., coronal heating) and to monitor solar variability for Sun-Earth connection studies. Each measurement of the solar spectrum sees a specific energy range that can either overlap with the energy ranges of other instruments or can leave gaps in the solar spectrum where no data are being taken. The Miniature X-ray Solar Spectrometer (MinXSS-1, \citealt{Mason2016}) mission and its much improved successor the Dual Aperture X-ray Solar Spectrometer (DAXSS) were designed to fill a previous void in observations of the solar spectrum between 0.5 and 10~keV. This energy range, in the soft X-ray (SXR) regime, is suitable for observing both the Bremsstrahlung continuum of the Sun as well as several major solar emission lines \citep{fletcher_2011}. For quiescent periods when the Sun's activity is relatively low, the spectral data can be used to study long-term changes of solar conditions such as average temperature and emission measure of the coronal plasma or active region evolution and development. During flaring times the SXR irradiance can increase by more than a factor of 100. With these data, studies can be performed on flare development, flare decay, and elemental abundance trends over the flaring period.

The MinXSS-1 CubeSat flew an X-ray spectrometer called an X-123 developed commercially from Amptek. Before flight, the X-123 was calibrated at the Synchrotron Ultraviolet Radiation Facility (SURF) at the National Institute for Standards and Technology (NIST) in Gaithersburg, Maryland \citep{surf_2011}. The calibration, discussed in detail in the paper \citep{moore2018}, found the detector's field-of-view (FOV) sensitivity, spectral resolution, spectral efficiency and effective area, detector response matrix, temperature response, and linearity of response. MinXSS-1 data are spread out around the solar minimum tail end of solar cycle 24, from 2016 June to 2017 April. Special interest was put into downlinking data around solar flares, however much of the data are representative of quiescent solar conditions especially when considering daily averages. The actual amount of data downlinked was around 1 percent of the total data acquired on orbit \citep{Mason2019}. This is due in part by having only one ground station at the time, enabling only a handful of passes per day, but also due to the method of data downlink being UHF, which has a lower bit rate than available on S-band or X-band. To fill in some of the missing data we have created a model of the SXR spectral irradiance using a ground-based solar proxy of 10.7~cm radio flux (F10.7).

The main processes that affect the variability of both 10.7~cm and extreme ultraviolet (EUV) emission occur in the same coronal plasma \citep{swarup1963high}. According to Kundu~1965, there are only two mechanisms that are sources of microwave radio emission from the Sun under non-flaring conditions \citep{kundu1965solar}. The first is thermal Bremsstrahlung, which is produced by free-free interactions between electrons and ionized particles \citep{wild1963solar}. The second is gyroresonance emission, which occurs via the acceleration of electrons spiraling around magnetic field lines \citep{tapping1987recent, schmahl1995microwave, schonfeld2015coronal}. Schonfeld~2015 explains that despite these two sources from the Sun, "Bremsstrahlung emission is closely related to EUV emission, while gyroresonance emission is not."

The correlation of F10.7 flux to solar activity level and the transparency of the atmosphere to 10.7~cm light has led to its use as a proxy measurement for solar EUV irradiance, which is highly attenuated by the Earth's atmosphere \citep{tobiska2008development, dudok2009finding, schonfeld2015coronal}. Since F10.7 value is widely accepted to indicate solar conditions it should be a good proxy data source to fill in gaps in data for the MinXSS-1 mission and others if a robust model can be made that correlates the two data sets. There are several previous models that use the F10.7 flux value to estimate solar variability, such as the Torr \& Torr~1979 model \citep{torr1979ionization}, the Hinteregger~1981 model \citep{hinteregger1981representations}, the EUVAC and HEUVAC models by Richards~1994 \& 2006 \citep{richards1994euvac, richards2006heuvac}, Solar2000/SIP by Kent Tobiska \citep{tobiska2008development}, and FISM/FISM2 by Phil Chamberlin \citep{chamberlin2008flare}. Each of these models have advantages over ours when modeling longer wavelengths or wider energy ranges, but they do not have the wavelength resolution down at the shorter SXR wavelengths to do a direct comparison to ours. Furthermore, these other proxy models are limited to the spectral resolution of the instrument that made the measurement, whereas the SWM model is not limited in this way.

Using MinXSS-1 data, a two temperature, two emission measure (2T2EM) model for the solar corona is made (Sec.~\ref{sec:2t2emmodel}). Then, in Sec.~\ref{sec:f10correlation}, the temperatures and emission measures found from the 2T2EM model are correlated to the F10.7 flux. In Sec.~\ref{sec:recreatingspectra} this correlation is used to recreate solar spectra from the F10.7 flux value alone, which are then plotted along with the original MinXSS-1 spectra. The generated model spectra are validated in Sec.~\ref{sec:goesvalidation} by calculating the daily irradiance from the model and comparing with the measured daily irradiance from GOES XRS-B. In Sec.~\ref{sec:fism2comparison} the irradiance between 1 to 8~Angstroms from the FISM-2 model are compared with that from the SWM model, and additional comparisons to other models (e.g. XPS DEM) are made in Sec.~\ref{sec:othercomparisons}. Finally, the conclusions are discussed in Sec.~\ref{sec:conclusion}.

\section{Data Products}
The first data product used in this analysis is the MinXSS-1 data that may be found online at lasp.colorado.edu/home/minxss/data/. The level 3 dataset is comprised of daily averages for all packet variables and spectra. These daily averages include in the calculated spectra contributions from solar flares and active regions, and does not, therefore, only represent pure quiescent daily averages. This is important to note since this may affect the correlations to the proxy data used, which is representative of the quiescent daily average.

The data product that is used as a proxy data set is the NOAA-filtered measurement of F10.7 taken by Penticton \citep{datacenter2022penticton}. The measurement of F10.7 represents the total emissions of 10.7~cm light from any source on the solar disk and is a good candidate upon which to base a proxy model because it is widely used as an indication of solar activity \citep{tapping201310}. The NOAA-filtered version of the Penticton data includes only the nearest to noon measurement, unless a solar flare or large radio burst is detected, then a value from another time of day is used.

The Geostationary Operational Environmental Satellite (GOES) X-Ray Sensor (XRS) \citep{garcia1994temperature} is an instrument that was operating during the MinXSS-1 mission. The XRS-B instrument is an ionization cell with a filter in place to limit the incoming light to wavelengths between 1 and 8 Angstroms (1.55-12.40~keV). The NOAA GOES XRS-B original data product is used here and provides an irradiance value in units of $Wm^{-2}$ that is taken in one minute averages. Using these data as comparison for other models has been done before and shown good agreement \citep{schwab2020soft}, and so will be of use to validate the accuracy of our proxy model. For our purposes the data are averaged to provide a daily average irradiance value. Daily averaging of both the MinXSS and XRS data smooths the large impulsive spikes of solar flares to give a more consistent irradiance value that describes the evolution of solar conditions such as active regions (AR) and quiescent sun (QS), but does still include these flaring times into the daily averages. This is an important note since the F10.7 measurement used does not include flaring times in its reported measurements.

One of the data products from the second generation of the Flare Irradiance Spectral Model (FISM), called FISM2 \citep{chamberlin2020flare}, is used. It is one of the more recent models that accurately calculates spectral irradiance variations due to the solar cycle, solar rotation, and solar flares. The data product used in this analysis for validation purposes is daily solar irradiance values, from 1947 to present, spanning the wavelengths of 0.1–190~nm. FISM2 uses many different solar proxies to estimate the variability of the solar spectral irradiance.  To determine the proxy model parameters, linear relationships are derived between each proxy and  spectral irradiance data sets from multiple satellite spectrometers, each with their own spectral range. It does not yet incorporate MinXSS-1 or DAXSS data, however, so there are higher uncertainties in the higher energy (shorter wavelength) ranges from 0.5~keV [2.48~nm] to 10~keV [0.12~nm] that these two missions cover. FISM2 also has an improved spectral bin resolution over the fist generation, now allowing 0.1~nm bin resolution. This model is used to quantify the changes in the solar VUV irradiance directly affecting satellite drag, radio communications, as well as the accuracy in the Global Positioning System (GPS) \citep{chamberlin2008flare}.

The spectra from the next generation of MinXSS instruments, called DAXSS, are used for further validation of this model with a spectra from a different instrument. The data from DAXSS cover a broader range of energies than MinXSS-1 and have a better spectral resolution. These spectra are also taken at a different time than MinXSS-1, covering higher levels of solar activity during the rise of the current solar cycle towards solar maximum. These DAXSS data are found at lasp.colorado.edu/home/minxss/data/ and are updated periodically since DAXSS is currently on orbit and downlinking new data every day.

For comparison of the temperatures and emission measures found by this model to those of existing models, the XUV Photometer System (XPS) differential emission measure (DEM) models for both quiet sun (QS) and active regions (AR) are used \citep{woods2022solar}. These models report DEM curves of emission measure vs temperature for all solar plasma above about 1~MK to higher than 14~MK. These curves are calculated separately for contributions from a QS and an AR.

\section{Two Temperature, Two Emission Measure Model}\label{sec:2t2emmodel}
Modeling the solar soft X-ray (SXR) spectra of the corona with two temperatures has been done before \citep{caspi2015new,schwab2020soft} and has shown good agreement with measured spectra. This process uses the atomic database from CHIANTI Version 9 \citep{dere_1997,dere_2019} with ionization fractions from Mazzotta 1998 \citep{mazzotta1998ionization} to generate solar spectra that contains both the emission lines within the selected energy ranges and the background continuum. The CHIANTI program is time intensive in calculating the radiation for millions of ion emissions for each corona configuration (temperature, emission measure, abundances). 

The IDL function \texttt{f\textunderscore vth.pro}, written by Richard Schwartz \citep{schwartz2002rhessi}, saves a large amount of computation time by interpolating between a pre-computed sample space of CHIANTI solutions with various temperatures and emission measures. The \texttt{f\textunderscore vth.pro} function then outputs a model solar spectrum for a given energy range when provided a temperature, emission measure, and relative elemental abundance factor as input parameters. The relative elemental abundance factor (AF) used is a scaling factor that is multiplied by a standard (or custom) list of elemental abundances. For this model the Feldman Standard Extended Coronal (FSEC) elemental abundance values \citep{feldman1992,Landi2002} were used with a scaling factor of unity.

To achieve a two-temperature, two-emission measure (2T2EM) model the \texttt{f\textunderscore vth.pro} procedure is used to create two model spectra. These spectra are created with identical energy ranges and energy binning then are summed together to create a single spectrum. This single 2T2EM model spectrum is then compared to a target MinXSS-1 spectrum to check the wellness of fit between the two. The input parameters for each temperature and emission measure are iteratively adjusted until a least squares minimum $\chi^2$ value is found. This is done using the IDL procedure \texttt{mpfit.pro} \citep{markwardt2009non}, which uses a Levenberg-Marquardt technique to iteratively find the best fit values.

\section{Correlation to F10.7 Flux}\label{sec:f10correlation}
Once the temperature (T) and emission measure (EM) parameters that best fit each daily averaged MinXSS-1 spectra are found, using the 2T2EM model described in Sec.~\ref{sec:2t2emmodel}, they are plotted against the daily F10.7 flux values to create a linear correlation we call the Schwab Woods Mason (SWM) model. It is seen that the warmer component T and EM as well as the cooler component EM, but not the cooler component T, increase and decrease with rise and fall of F10.7 flux. The cooler component T stays almost constant for all levels of F10.7 flux and solar intensity. The T and EM parameters correlate to the F10.7 flux with a linear correlation. To obtain a calculation of the slope and offset of each linear relationship the program \texttt{robust\textunderscore linefit.pro} is used \citep{Freudenreich1991}. 

Initially the correlations had a number of outlier points that were not statistically valid. A basic linear fitting method is used first to obtain the linear relationship but due to the outlier points in the fit values could not get accurate results. This showed the necessity of using the more robust linear fitting program \texttt{robust\textunderscore linefit.pro}, which is more outlier-resistant \citep{Freudenreich1991}. To clean up the data, a filter is applied letting only the daily fits through where all four of its fit parameters were within one sigma of the robust linear fit line. This filtering technique did not change the trends of the data but eliminated the outlier points that were impairing the linear fits. Another robust linear fit is applied to the filtered data and the process is repeated until the slopes found between consecutive linear fits changed by less than 1 percent. The resulting linear fit values are shown in Table~\ref{table:meanandlinearfit}.

\begin{figure}[t]
    \centering
    \includegraphics[width=\linewidth]{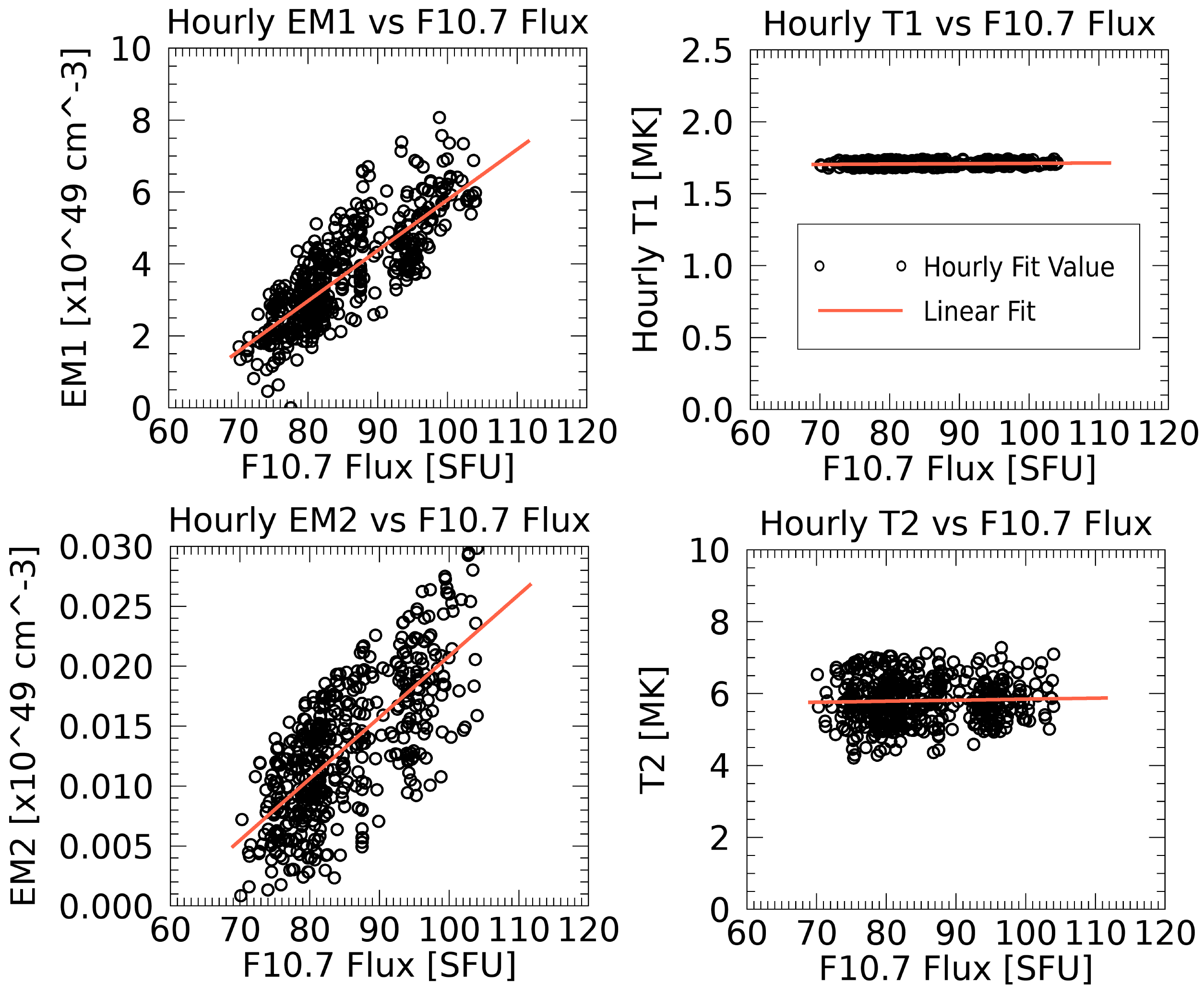}
    \caption{The hourly averaged MinXSS-1 data are fit to a two-temperature, two-emission measure model and are shown plotted against the interpolated F10.7 flux value of the same hour. Robust linear fits are applied to the filtered data, showing four separate linear relationships to F10.7 flux. \label{fig:hourlyf10}}
\end{figure}

\begin{figure}[b]
    \centering
    \includegraphics[width=\linewidth]{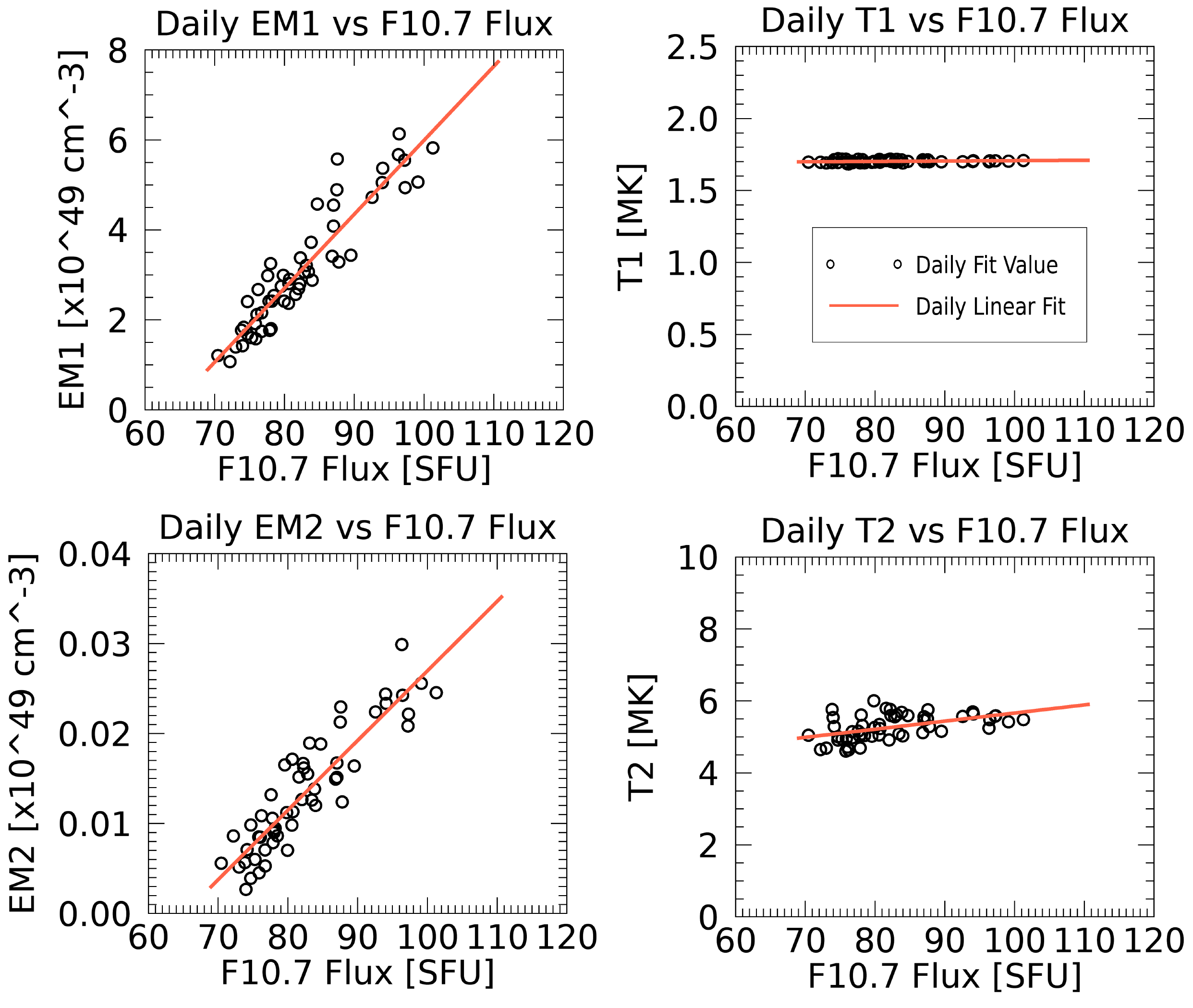}
    \caption{Same as Fig.~\ref{fig:hourlyf10} but for daily averaged MinXSS-1 data. \label{fig:dailyf10}}
\end{figure}

The SWM model of both daily and hourly averaged data both provide two EM correlations that increase as the F10.7 flux increases with a linear relationship. For the hourly SWM model shown in Fig.~\ref{fig:hourlyf10}, the two T correlations show that both T values remain almost constant as F10.7 flux increases, however for the daily SWM model shown in Fig.~\ref{fig:dailyf10} only the cooler T component shows a relationship that is almost independent of F10.7 flux value. The slope of the warmer temperature component had a non-zero slope that did vary with F10.7 flux and thus varies with solar intensity. The near-zero slope of the cooler temperature component suggests that although two temperatures can describe the daily average solar SXR, the temperature of the cooler component is essentially independent of solar activity and likely describes the quiet sun (QS). The warmer temperature component then describes contributions from any active regions (AR). The distinction between cool and warm T components representing the QS and AR contributions is discussed further in Sec.~\ref{sec:othercomparisons}. 

\begin{deluxetable*}{l c c c c}[!h]
\tablecaption{SWM Model Mean Values and Linear Correlations to F10.7 Flux \label{table:meanandlinearfit}}
\tabletypesize{\scriptsize}
\tablehead{\colhead{} & \colhead{EM1} & \colhead{T1} & \colhead{EM2} & \colhead{T2} \\
 \colhead{} & \colhead{($\num{e49}\ cm^{-3}$)} & \colhead{(MK)} & \colhead{($\num{e49}\ cm^{-3}$)} & \colhead{(MK)}
} 
\startdata
    {Daily Mean} & 3.14 $\pm$ 0.48 & 1.703 $\pm$ 0.009 & 0.014 $\pm$ 0.003 & 5.27 $\pm$ 0.31 \\ 
    {Hourly Mean} & 3.67 $\pm$ 0.81 & 1.707 $\pm$ 0.015 & 0.013 $\pm$ 0.004 & 5.80 $\pm$ 0.63 \\
    {Daily Slope} & 0.16 $\pm$ 0.01 & (2.4 $\pm$ 1.0)$\times10^{-4}$ & (7.7 $\pm$ 0.5)$\times10^{-4}$ & (2.3 $\pm$ 0.5) $\times10^{-2}$ \\ 
    {Hourly Slope} & 0.14 $\pm$ 0.01 & (2.7 $\pm$ 0.8)$\times10^{-4}$ & (5.1 $\pm$ 0.2)$\times10^{-4}$  & (0.3 $\pm$ 0.4) $\times10^{-2}$ \\
    {Daily Offset*} & 1.06 $\pm$ 0.70 & 1.700 $\pm$ 0.01 & 0.004 $\pm$ 0.004 & 4.99 $\pm$ 0.4 \\ 
    {Hourly Offset*} & 1.57 $\pm$ 0.39 & 1.703 $\pm$ 0.01 & 0.005 $\pm$ 0.002 & 5.76 $\pm$ 0.3 \\
\enddata
\tablecomments{The offset value is found by setting F10.7 Flux to a value of 70 instead of 0. The slope values are per SFU. An abundance factor of unity is used, meaning the elemental coronal abundances are the Feldman Standard Extended Coronal values \citep{feldman1992,Landi2002}}
\end{deluxetable*}

\begin{figure*}[h]
    \centering
    \includegraphics[width=\textwidth]{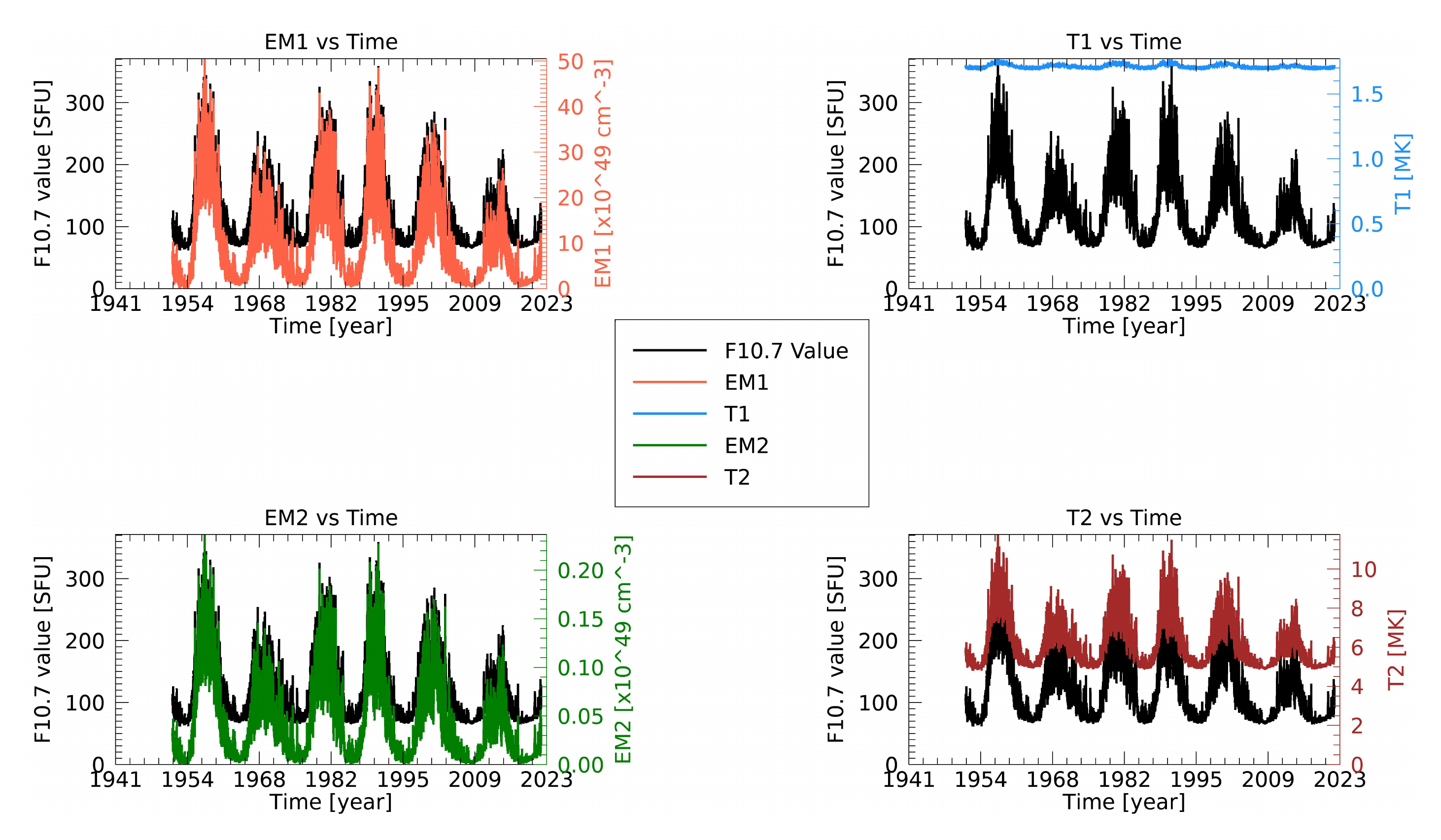}
    \caption{The evolution of the two temperatures and two emission measures that describe the daily average solar coronal plasma conditions are shown to increase and decrease along with the F10.7 flux value over several solar cycles. \label{fig:f10historicalcorrelations}}
\end{figure*}

The cooler temperature component of 1.7~MK is consistent for both hourly and daily linear correlations to F10.7 flux. The warmer temperature component ranges between 5~MK to 6~MK for daily averaged data and a similar temperature range across all levels of solar activity, having a mean of 5.8~MK for hourly fits. To account for the increase or decrease in solar intensity from day to day the EM attributed to each T is scaled based on the linear relationships in Table~\ref{table:meanandlinearfit}. The daily SWM model linear correlations were chosen over the hourly correlations because it is found that the accuracy of the daily model is better over a wider range of F10.7 values than the hourly model. The higher uncertainties for the hourly comparisons are somewhat expected because the MinXSS hourly averages can have larger flare contributions and because the F10.7 is a just a daily proxy.

Since F10.7 flux measurements date back to 1947, this model may be applied to provide the two T and two EM values that show coronal variation over more than six solar cycles. For each day with F10.7 flux equal to F, first subtract 70 from F to get (F-70) then multiply this number against the slope value for that parameter listed in Table~\ref{table:meanandlinearfit} and finally add its offset. The two T and two EM plots since 1947 are shown in Fig.~\ref{fig:f10historicalcorrelations} and depict the long term temperature and density evolution of the corona. To obtain these values of T and EM the daily slope and offset correlations are used from the SWM model. The trend of the cooler temperature component shows that despite the fluctuations of F10.7 flux over a solar cycle, the T of the cooler component remains almost unchanged. Conversely, the T of the warmer component is dependent on the levels of F10.7 flux and solar activity. The associated EM of both the cooler and warmer components follows F10.7 flux and accounts for the change in solar intensity. This model suggests that during times of low to moderate solar activity the warmer temperature component varies between about 5 to 6~MK, but during times of moderate to high solar activity this component can get higher than 10~MK. These higher temperatures ($\ge10~MK$) have been shown in some DEM analyses such as \citep{caspi2015new, woods2022solar} and are compared later in Sec.~\ref{sec:othercomparisons}. However, this model is generated using MinXSS-1 data, which was only taken during times of low to moderate solar activity. Additional measurements from a broader range of times would provide better constraints on the relationships of F10.7 during times of higher solar activity.

\section{Recreating Solar Spectra from F10.7 Flux}\label{sec:recreatingspectra}
Using the linear correlations described in Sec.~\ref{sec:f10correlation} and shown in Table~\ref{table:meanandlinearfit} we find for each day the two temperatures (2T) and two emission measures (2EM) that describe the daily average solar corona. These daily 2T and 2EM values are derived from the daily F10.7 flux alone. Passing these daily values into the \texttt{f\textunderscore vth.pro} procedure discussed in Sec.~\ref{sec:2t2emmodel} the model can recreate SXR spectra for a given energy range. This energy range is not restricted by the observable energy range of the instrument and is only restricted by accuracy of this two temperature model describing a spectrum over energies outside of 1~keV to 10~keV. The accuracy beyond this region is not fully tested or known. Another key benefit of this model is that it is not limited in spectral resolution since it does not depend on the resolution of any instrument, only the daily F10.7 flux value, and therefore allows for spectra to be generated at any spectral resolution.

\begin{figure}[H]
    \centering
    \includegraphics[width=\linewidth]{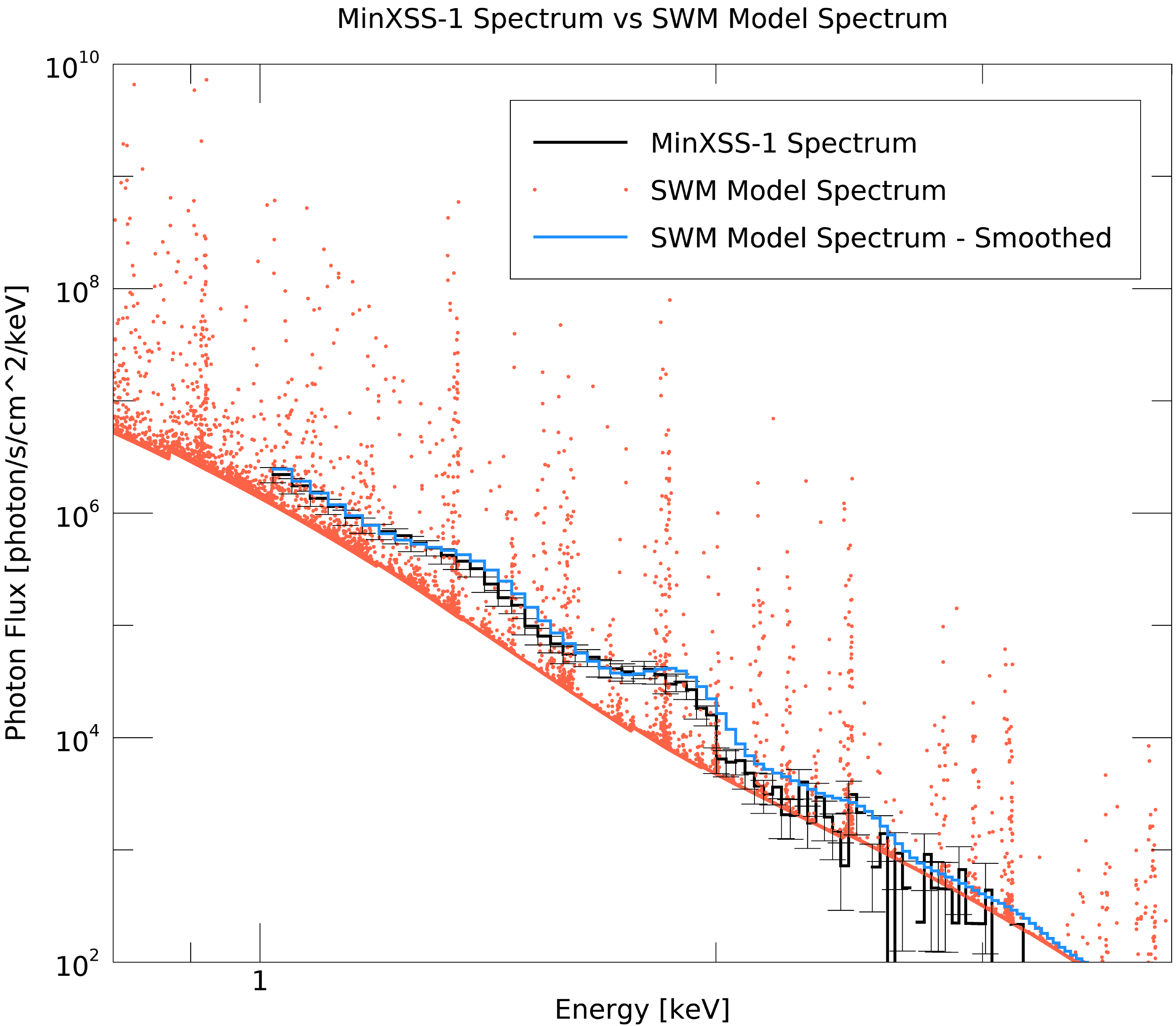}
    \caption{\label{fig:minxss1validation} Comparison between the F10.7 flux model spectrum and the corresponding MinXSS-1 daily averaged spectrum (shown in black). The SWM model spectrum is shown in red points and the spectrum smoothed with the resolution and response matrix of MinXSS-1 is shown in blue. The F10.7 value on this day was 76.7 SFU.}
\end{figure}

A comparison between the F10.7 flux model spectrum and the corresponding MinXSS-1 daily averaged spectrum is shown in Fig.~\ref{fig:minxss1validation}. The modeled spectrum is shown in red points. The resolution of the raw model spectra or number of points to use is arbitrary, but one million points over the energy range 0.01~keV to 30~keV is plotted here. The raw model spectrum is smoothed to the resolution of the MinXSS-1 X123 spectrometer to give the smoothed model spectrum that is shown in blue. This smoothing process takes into account the full width half maximum resolution of the detector and response matrix of each energy bin relating counts per second on the detector to photon flux. More information about this smoothing process can be found in \citep{schwab2020soft}. 

The smoothed SWM model spectrum and the measured daily average MinXSS-1 spectrum for the same day show good agreement with one another. The energy range for comparison between model and measured spectra is limited to energies above 1~keV since the MinXSS-1 detector is well calibrated for measurements above this energy, but has some low energy noise below about 0.8~keV that is not entirely accounted for in the response matrix. It can be seen here that for days with low solar activity, such as the day this spectrum was taken, the MinXSS-1 spectra become noisy around 2.0~keV due to the low amount of detectable photons in the higher energies. The counting statistics are too low to resolve a spectrum above around 2.0~keV during quiescent times. 

With this model MinXSS-1 spectra is recreated from the daily F10.7 flux measurement without the restriction of low photon detection noise in the higher energies. These higher energy model spectra may still be smoothed to the MinXSS-1 detector resolution and response since it is well calibrated from about 1 to 10~keV. For energies lower than 1~keV there is noise that is unaccounted for so a smoothed model may not be accurate, however a model may still be made for these energies since the model is not limited to the response or resolution of any detectors. Furthermore, the temperatures and emission measures may be used in any other software that generates solar spectra from temperature and emission measure.

\section{Validation with GOES XRS-B Irradiance}\label{sec:goesvalidation}

As a validation of the SWM model and the spectra that it generates, the model's daily irradiance between 1 to 8 Angstroms is compared with the daily GOES XRS-B \citep{garcia1994temperature} measured irradiance values of the same wavelength regime. The data points shown span the dates 1996 January 01 to 2018 April 30 since the daily cadence was very reliable past 1996 and NOAA stopped updating their historical dataset past 2018 April 30. This subset data, from 1996 to 2018, spans about two full solar cycles including both solar minima and solar maxima.

The MinXSS spectra are converted from photon flux in units of $[{ph}\ cm^{-2}s^{-1}keV^{-1}]$ to irradiance between 1 to 8~Angstroms in units of $[{W}m^{-2}]$ by summing the contributions of each energy bin as shown in Eq~\ref{eqn:irradiance}. 

\begin{equation}\label{eqn:irradiance}
    Irr=\sum_{E=1.5498}^{12.3984}\left[F*E\right]dE\cdot(\num{1.6022e-16})\cdot(\num{1e4})
\end{equation}
where $Irr$ is the irradiance in units of $[{W}m^{-2}]$, $F$ is the photon flux in units of $[{ph}\ cm^{-2}s^{-1}keV^{-1}]$, $E$ is the energy of each bin in $[keV]$, $dE$ is the energy bin width of $0.02975~[keV]$, $\num{1.6022e-16}~[J*keV^{-1}]$ is the conversion factor between keV and Joules, and $\num{1e4}~[cm^{2}m^{-2}]$ is the conversion factor between $cm^{-2}$ and $m^{-2}$.

\begin{figure*}[h]
    \centering
    \includegraphics[width=\textwidth]{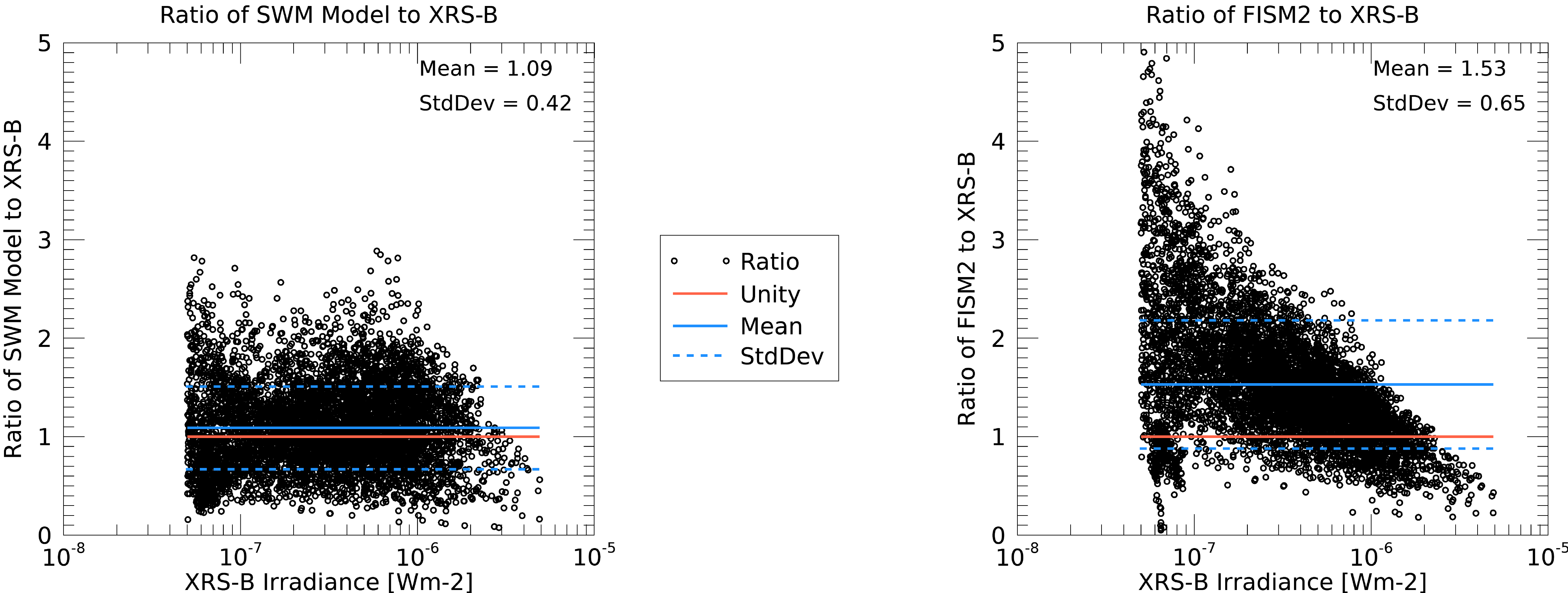}
    \caption{Left:} The irradiances of the modeled spectra between 1-8~Angstroms are compared with the measured irradiances of the GOES XRS-B instrument of the same wavelength regime. On average the ratio between the modeled irradiance and the XRS-B irradiance is close to unity, with a mean value of 1.09 and a standard deviation of 0.42. Right: The irradiances generated from FISM2 between 1-8~Angstroms are compared with the measured irradiances of the GOES XRS-B instrument of the same wavelength regime. For low levels of XRS-B irradiance FISM2 calculates an irradiance value that is as much as a factor of 6 higher than the measured irradiance from XRS-B, but for levels around $\num{e-6}$~Wm-2 the FISM2 model and our model agree with one another as well as agree with XRS measurements on average. \label{fig:FISMandGOESratios}
\end{figure*}

After the calculation of model irradiance between 1 to 8~Angstroms has been made it is compared with the measured average daily irradiance value from GOES XRS-B. A direct comparison of modeled irradiance to measured irradiance from XRS is done and a linear fit between them is calculated. The ratio mean is found to be 1.09, which shows agreement between the model and measured values. The left-side plot of Fig.~\ref{fig:FISMandGOESratios} is a different representation of the data that shows the ratio of the model irradiance to the measured irradiance from XRS-B. It can be seen that although the ratio varies for all levels of solar activity, on average the ratio is slightly above, but close to unity. For high solar irradiance the ratio tends to underestimate the measured irradiance, indicating that on average the model is best for middle to low levels of solar irradiance. The total measured F10.7 emission is a sum of several different sources; the main two being Bremsstrahlung and gyroresonance \citep{kundu1965solar}. The underestimation of F10.7 for higher levels of solar activity may be explained in part by the uneven contributions from these different components to the F10.7 emission. Contributions from gyroresonance emission increase with the high magnetic fields in active regions \citep{tapping201310}, so this contribution becomes more relevant during times with higher solar activity. The SWM model is an empirical model that is trying to capture the average behavior of all F10.7 emission, meaning this could explain the non-linearity during times of higher solar activity. This model also is currently only using data from MinXSS-1 that was taken at low to moderate activity. It is planned to address this once the DAXSS instrument, currently on orbit and taking measurements, has sufficient observations during times of moderate to high solar activity.

The mean value of the ratio between the modeled irradiance and the measured value is 1.09 with a standard deviation of 0.42. There are two main reasons that this standard deviation is not lower. The first is that flaring data are not represented in the F10.7 measurements but are included in XRS and MinXSS data. The second is that the hotter temperature is looking at a mean temperature of AR contributions instead of individual AR contributions. There are days with multiple ARs on the solar disk that each has a different characteristic temperature and emission measure and therefore not exactly representative of a single AR contribution. Also, multiple ARs could be at different stages of their own evolution.


\section{Comparison to FISM2}\label{sec:fism2comparison}

To compare the accuracy and utility of this model to others in existence, the irradiance calculations between 1-8~Angstroms are compared to those from FISM2. The FISM2 daily average data are available on the Laboratory for Atmospheric and Space Physics online LISIRD data center (lasp.colorado.edu/lisird/) and the modeled irradiances of 7 wavelength bins of width 0.1~nm (centered on 0.15~nm, 0.25~nm, ..., 0.75~nm) were downloaded and summed together to get the total irradiance from 1-8~Angstroms. The irradiance values from FISM2 are in units of $[W/m^2/nm]$, so these values are multiplied by the bin width of 0.1~nm to convert to irradiance in units of $[W/m^2]$ that can be directly compared to the 1-8~Angstrom irradiance calculated from our model and to that measured from GOES XRS-B.

Ideally if FISM2 agreed perfectly with our model there would be a ratio equal to unity between the two irradiance calculations. It is shown from the left side plot in Fig.~\ref{fig:FISMandGOESratios} that for XRS-B irradiance levels around $\num{e-6}$~$Wm^-2$ the models agree with one another and from Sec.~\ref{sec:goesvalidation} we have shown that our model agrees with the true measured values on average. For lower irradiance values there is a larger discrepancy of FISM2 to both our model and to the measured irradiance. This shows that FISM2 has higher uncertainties than our model between 1-8~Angstroms and that for modeling irradiance in this range the SWM model is more accurate and precise. This is likely due to the current absence of instrument spectral data in this regime, such as MinXSS-1, being incorporated into the FISM-2 model.

\section{Validation with DAXSS Data}\label{sec:daxssvalidation}
The INSPIRESat-1 CubeSat was recently launched on 2022 February 14 and had onboard an improved version of the X123 X-ray spectrometer that flew on MinXSS-1 \citep{chandran2021inspiresat}. The instrument onboard is the Dual Aperture X-ray Solar Spectrometer (DAXSS) that has improvements both in the design of the spectrometer itself as well as in the design of its dual aperture \citep{schwab2020soft}.

\begin{figure}[H]
    \centering
    \includegraphics[width=\linewidth]{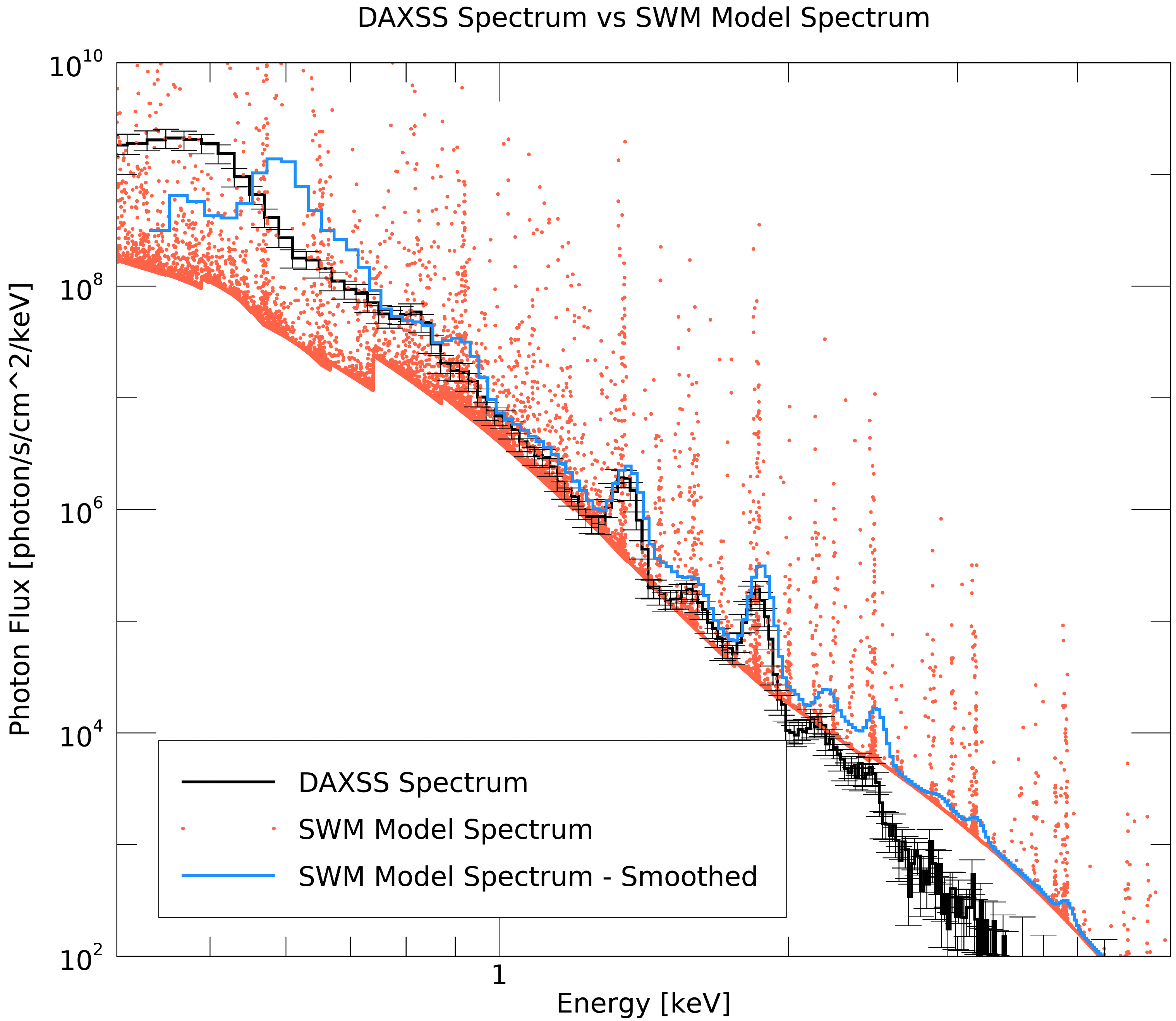}
    \caption{\label{fig:daxssvalidation}Comparison between the F10.7 flux model spectrum and the DAXSS spectrum of the same day (shown in black). The SWM model spectrum is shown in red points and the spectrum smoothed with the resolution and response matrix of DAXSS is shown in blue.}
\end{figure}

Spectra are recreated from the F10.7 flux value alone as described in Sec.~\ref{sec:recreatingspectra}. Since DAXSS is a newer mission and NOAA has stopped updating their historical F10.7 dataset as of April 30, 2018 the F10.7 value is taken instead from the CLS Solar Radio Flux at 10.7 cm dataset that can also be found in the LISIRD system \citep{ware2010solar}. Since the current DAXSS data are taken when the solar cycle is approaching solar maximum, the F10.7 flux values are on average higher than those during the MinXSS-1 mission. The plot shown in Fig.~\ref{fig:daxssvalidation} is for a day where the F10.7 flux value was 108.7 SFU. This figure shows very good agreement between the model and the DAXSS spectrum  for energies between about 0.7~keV and 2.5~keV but shows divergence outside this energy range. The instrument response of DAXSS is not well calibrated below 0.7~keV, so the disagreement below 0.7~keV likely means that the measured spectra need to be adjusted. This shows a perk of having a model that can extend to lower energies than the instrument.

\section{Comparison to Other Models}\label{sec:othercomparisons}
Temperature fitting has been looked into for quiescent sun (QS), active regions (AR), and flares. Some models are done using only one temperature, some with two temperatures, and some are differential emission measure (DEM) models with many temperatures. For each model there are some advantages of simplicity and other advantages for added complexity. For example, the temperature of the quiescent solar corona is between 1~MK to 2~MK \citep{sakurai2017heating} and so a quiescent model may be made that has a single temperature component of 1~MK to 2~MK. To add more more complexity, we can consider the temperature of an active region on the Sun. Active regions span a wider range of energies for an isothermal model that are typically between 3~MK to 10~MK \citep{yoshida1996temperature}.

Whole sun models are those that do not zoom in on particular regions of the sun, for example only on the active region. Including two temperatures in a whole sun model can account for both the contribution of the QS and any active regions present. For a two-temperature (2T) model the cooler temperature (T1) component usually accounts for the QS contributions and the hotter temperature (T2) component usually account for any AR contributions. During B3 levels of solar activity some 2T models have shown that T1 is around 1.7 MK to 1.9 MK and T2 is around 4.5 MK to 5.3 MK \citep{moore2018}.

\begin{figure}[H]
    \centering
    \includegraphics[width=\linewidth]{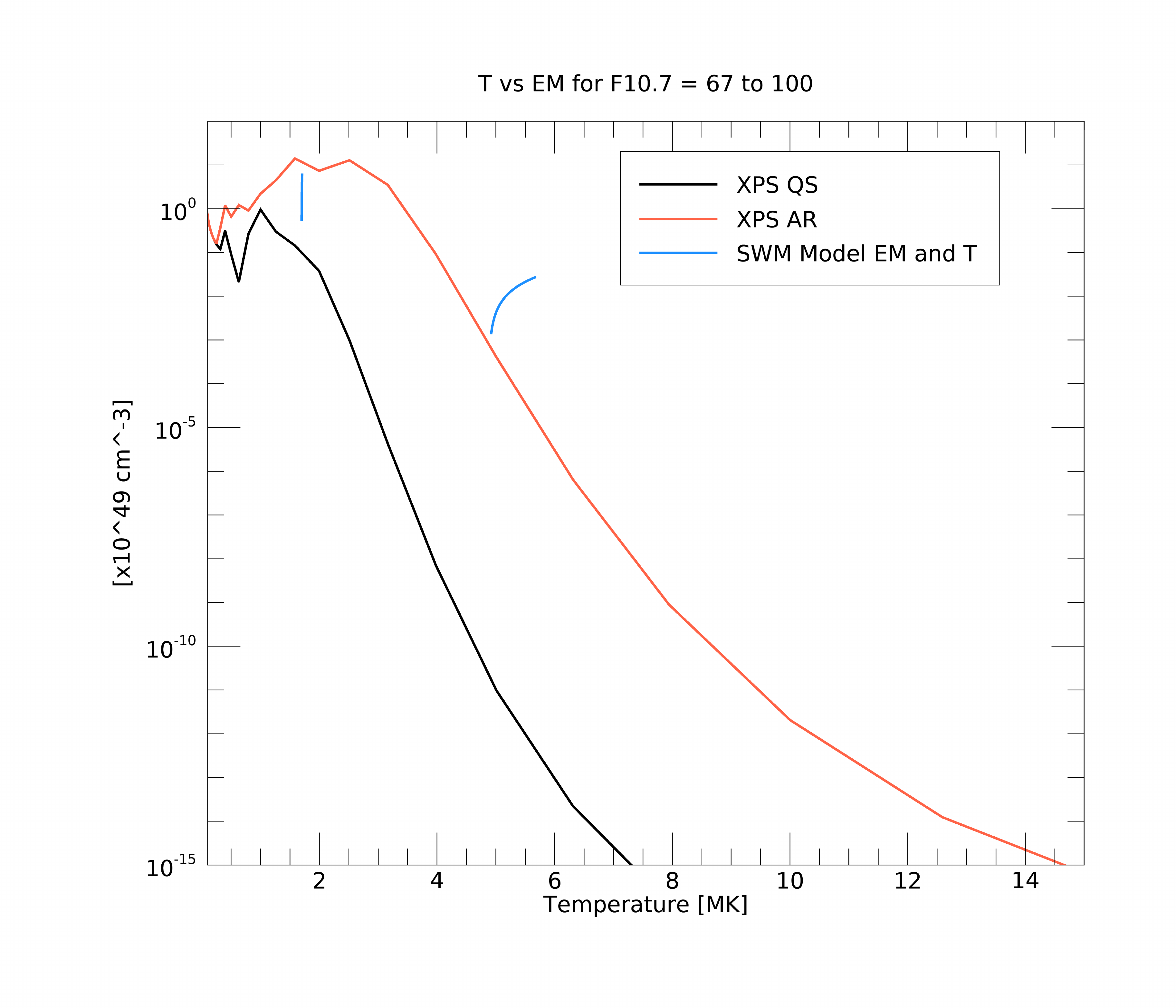}
    \caption{\label{fig:xpsdemcomparison}Comparison between the two temperatures and two emission measures of the SWM model and the XPS differential emission measure models for quiescent sun (QS) and active region (AR) X-ray flux contributions for low solar intensity of F10.7 value = 67~SFU to higher solar intensity of F10.7 value = 100~SFU.}
\end{figure}

DEM models have the advantage of including many more temperature components each with its own emission measure. This level of added complexity shows that the solar corona is not only two temperatures, but has a large range of temperatures. Some DEM curves are made to model specific spectra such as in \citep{caspi2015new} that show a model with temperatures in the range of 1.8~MK to 10.7~MK. Other DEM models, such as the XUV Photometer System (XPS) DEM are made to be representations of average contributions of QS and AR \citep{woods2022solar}. Fig.~\ref{fig:xpsdemcomparison} shows the comparison between the SWM model with a F10.7 value ranging from 67~SFU to 100~SFU and the XPS DEM models for QS and AR. It can be seen here that T1 aligns closely with the QS DEM curve and T2 aligns more closely with the AR DEM curve, especially at lower solar intensity. Both components rise above their respective curves and this is due to the contributions of the SWM model being only from two temperatures and two emission measures instead of spread out over many temperatures and emission measures.

\section{Conclusions}\label{sec:conclusion}
A two-temperature, two-emission measure (2T2EM) model is made for each daily averaged and hourly averaged MinXSS-1 spectrum. Each of the temperature (T) and emission measure (EM) parameters found in the 2T2EM model fits are then correlated to the F10.7 flux value when the spectrum was taken. Four linear relationships that make up the Schwab Woods Mason (SWM) model are found between each parameter and the F10.7 flux value, shown in Fig.~\ref{fig:hourlyf10} for hourly correlations and Fig.~\ref{fig:dailyf10} for daily correlations. The mean values as well as the linear fit slopes, offsets, and the uncertainties of each are shown in Table~\ref{table:meanandlinearfit}. The accuracy of the daily SWM model is found to be better over a wider range of F10.7 values than the hourly model. This is likely due to the interpolation of F10.7 values onto an hourly scale failing to vary with hourly variations on the Sun such as active region development or solar flares. The cooler temperature component of the daily SWM model is essentially independent of F10.7 flux as well as solar intensity, which is shown in Fig.~\ref{fig:f10historicalcorrelations} over more than six solar cycles since 1947. The T of the cooler component remains almost constant at a T of 1.7~MK and its corresponding EM accounts for the varying levels of solar intensity. The T of the warmer component varies over the solar cycles between 5~MK and 6~MK depending on solar activity, and its corresponding EM increases or decreases to account for higher or lower levels of solar intensity as well.

MinXSS-1 spectra are recreated using only the F10.7 value from the same day, as shown in Fig.~\ref{fig:minxss1validation}. The \texttt{f\textunderscore vth.pro} procedure is used in the spectrum recreation process with inputs of two T and two EM found from the linear correlations. The model spectra may be smoothed to match the energy resolution of MinXSS-1 using its response function. Smoothed model spectra and measured MinXSS-1 spectra agree very nicely for energies above 1~keV and the model is able to extend into higher or lower energies than observable with MinXSS-1. The extremely high spectral resolution of this model is also shown in Fig.~\ref{fig:minxss1validation}, making this model more resolved over a wider energy range than previous and even current proxy models in the SXR regime.

Model spectra are created for each day from 1996 January 01 to 2018 April 30. The irradiance between 1 and 8~Angstroms of each spectrum is calculated and compared to the measured GOES XRS-B irradiance of the same wavelength regime. A one-to-one comparison between the model and measured irradiances is done and a linear fit is calculated to this comparison. Fig.~\ref{fig:FISMandGOESratios} shows through a linear fit that the ratio mean is 1.09 and compares well to an ideal ratio of unity on average. The ratio plot also shows that for higher levels of solar activity, above about a GOES C2 level, the modeled irradiance tends to be smaller than the measured value.

The comparison between our model and FISM2 shows that FISM2 has higher uncertainties than our model between 1-8~Angstroms and that for modeling irradiance between 0.5-10~keV our model is more accurate and precise. With FISM2 it is possible to model a wider region, from 0.1-190~nm, at a spectral bin size of 0.1~nm, but our model has better spectral resolution within the modeled region that is only limited by the allowed resolution of CHIANTI. FISM2 does not model low irradiances well inside the MinXSS-1 energy range, however our model lacks accuracy for irradiances higher than $2e{-6}$ $W/m^2$, which is when F10.7 flux is higher than about 100~SFU. Another benefit of the SWM model compared with FISM2 is that in addition to calculating irradiance values, as well as daily averaged spectra, the SWM model provides physical values of temperature and emission measure from the 2T2EM model for each day. These improvements of our model indicates that FISM would benefit from this model as well as including a similar model for DAXSS. The work to incorporate this model into FISM3 has already begun. 

To further validate this model, spectra are generated for the DAXSS instrument that is currently on orbit. Fig.~\ref{fig:daxssvalidation} shows very good agreement between the model and the DAXSS spectrum for energies below about 2.3~keV but a slight divergence above this energy. This could be attributed to either the fact that the F10.7 flux on that day was 108.7 SFU and is at the higher end of the range that the MinXSS-1 mission covered or possibly due to MinXSS-1 energy range being smaller than that of DAXSS. To address this discrepancy, a further study looking into the relationship between DAXSS data (with higher-resolution and wider-energy range) to F10.7 flux should be done.

Lastly, the SWM model is compared to other temperature models of the Sun. The lower of the two temperatures in the 2T2EM model is more representative of the quiescent sun while the higher temperature is more representative of any active regions that may be present. This can be seen in Fig.~\ref{fig:xpsdemcomparison} in the comparison between XPS DEM curves for QS and AR and the T and EM values for when F10.7 is at a low intensity value of 67 SFU.

In conclusion, the model described in this paper is a useful tool for modeling the temperatures and emission measures that describe the solar corona on a day-to-day cadence that may be used to study the long-term evolution of the corona. This model also can generate SXR spectra and irradiances that cover a wider energy range and temporal range (back to 1947) than was observed by the original MinXSS-1 spectrometer, as well as fill in missing spectra. The spectra that this model generate are also more resolved over a wider energy range than previous and even current proxy models in the SXR regime. These improvements in modeling will be absorbed into the next generation FISM3.

\subsection{Acknowledgements}
This research was performed with funding from the NASA Grant NNX17AI71G to the University of Colorado (CU). We are immensely thankful for the NIST SURF staff for their excellent support of calibrations at their facilities in Gaithersburg, MD. A special thanks to Mitch Furst for his help while at NIST SURF. We would like to acknowledge the work done by Richard Schwartz at NASA to develop the \texttt{f\textunderscore vth.pro} procedure in IDL that was used in our model fits. The majority of this research was done by CU graduate student Bennet Schwab, and he is vastly grateful to CU Professor Scott Palo and Dr. Tom Woods for their graduate academic advice and scientific guidance.

\software{
	AASTeX \citep{AASJournalsTeam2018}, 
	IDL \citep{schwartz2002rhessi},
	OSPEX software \citep{tolbert2020ospex},
	SolarSoft \citep{SolarSoft2012},
	}

\bibliography{F10flux_Paper_Bibliography}{}

\begin{thebibliography}{}
\expandafter\ifx\csname natexlab\endcsname\relax\def\natexlab#1{#1}\fi
\providecommand{\url}[1]{\href{#1}{#1}}
\providecommand{\dodoi}[1]{doi:~\href{http://doi.org/#1}{\nolinkurl{#1}}}
\providecommand{\doeprint}[1]{\href{http://ascl.net/#1}{\nolinkurl{http://ascl.net/#1}}}
\providecommand{\doarXiv}[1]{\href{https://arxiv.org/abs/#1}{\nolinkurl{https://arxiv.org/abs/#1}}}

\bibitem[{{AAS Journals Team} \& Hendrickson(2018)}]{AASJournalsTeam2018}
{AAS Journals Team}, \& Hendrickson, A. 2018, {AASJournals/AASTeX60: Version
  6.2 official release}, \dodoi{10.5281/ZENODO.1209290}

\bibitem[{Arp {et~al.}(2011)Arp, Clark, Deng, Faradzhev, Farrell, Furst,
  Grantham, Hagley, Hill, Lucatorto, \& et~al.}]{surf_2011}
Arp, U., Clark, C., Deng, L., {et~al.} 2011, Nuclear Instruments and Methods in
  Physics Research Section A: Accelerators, Spectrometers, Detectors and
  Associated Equipment, 649, 12–14, \dodoi{10.1016/j.nima.2010.11.078}

\bibitem[{Caspi {et~al.}(2015)Caspi, Woods, \& Warren}]{caspi2015new}
Caspi, A., Woods, T.~N., \& Warren, H.~P. 2015, The Astrophysical Journal
  Letters, 802, L2

\bibitem[{Chamberlin {et~al.}(2008)Chamberlin, Woods, \&
  Eparvier}]{chamberlin2008flare}
Chamberlin, P.~C., Woods, T.~N., \& Eparvier, F.~G. 2008, Space Weather, 6

\bibitem[{Chamberlin {et~al.}(2020)Chamberlin, Eparvier, Knoer, Leise,
  Pankratz, Snow, Templeman, Thiemann, Woodraska, \&
  Woods}]{chamberlin2020flare}
Chamberlin, P.~C., Eparvier, F.~G., Knoer, V., {et~al.} 2020, Space Weather,
  18, e2020SW002588

\bibitem[{Chandran {et~al.}(2021)Chandran, Fang, Chang, Hari, Woods, Chao,
  Kohnert, Verma, Boyajian, Duann, {et~al.}}]{chandran2021inspiresat}
Chandran, A., Fang, T.-W., Chang, L., {et~al.} 2021, Advances in Space
  Research, 68, 2616

\bibitem[{Datacenter(2022)}]{datacenter2022penticton}
Datacenter, L. I. S.~I. 2022, Penticton Solar Radio Flux at 10.7 cm, Time
  Series [data set]

\bibitem[{Dere {et~al.}(2019)Dere, Del~Zanna, Young, Landi, \&
  Sutherland}]{dere_2019}
Dere, K.~P., Del~Zanna, G., Young, P.~R., Landi, E., \& Sutherland, R.~S. 2019,
  The Astrophysical Journal Supplement Series, 241, 9,
  \dodoi{https://doi.org/10.3847/1538-4365/ab05cf}

\bibitem[{Dere {et~al.}(1997)Dere, Landi, Mason, Fossi, \& Young}]{dere_1997}
Dere, K.~P., Landi, E., Mason, H.~E., Fossi, B. C.~M., \& Young, P.~R. 1997,
  Astronomy and Astrophysics Supplement Series, 125, 149–173,
  \dodoi{10.1051/aas:1997368}

\bibitem[{Dudok~de Wit {et~al.}(2009)Dudok~de Wit, Kretzschmar, Lilensten, \&
  Woods}]{dudok2009finding}
Dudok~de Wit, T., Kretzschmar, M., Lilensten, J., \& Woods, T. 2009,
  Geophysical Research Letters, 36

\bibitem[{{Feldman} {et~al.}(1992){Feldman}, {Mandelbaum}, {Seely}, {Doschek},
  \& {Gursky}}]{feldman1992}
{Feldman}, U., {Mandelbaum}, P., {Seely}, J.~F., {Doschek}, G.~A., \& {Gursky},
  H. 1992, \apjs, 81, 387, \dodoi{10.1086/191698}

\bibitem[{Fletcher {et~al.}(2011)Fletcher, Dennis, Hudson, Krucker, Phillips,
  Veronig, Battaglia, Bone, Caspi, Chen, Gallagher, Grigis, Ji, Liu, Milligan,
  \& Temmer}]{fletcher_2011}
Fletcher, L., Dennis, B.~R., Hudson, H.~S., {et~al.} 2011, Space Science
  Reviews, 159, 19, \dodoi{10.1007/s11214-010-9701-8}

\bibitem[{Freeland \& Handy(2012)}]{SolarSoft2012}
Freeland, S., \& Handy, B. 2012, {SolarSoft: Programming and data analysis
  environment for solar physics}

\bibitem[{Freudenreich(1991)}]{Freudenreich1991}
Freudenreich, H. 1991, Robust Linefit.pro,
  \url{https://idlastro.gsfc.nasa.gov/ftp/pro/robust/robust_linefit.pro}

\bibitem[{Garcia(1994)}]{garcia1994temperature}
Garcia, H.~A. 1994, Solar Physics, 154, 275

\bibitem[{Hinteregger(1981)}]{hinteregger1981representations}
Hinteregger, H. 1981, Advances in Space Research, 1, 39

\bibitem[{Kundu(1965)}]{kundu1965solar}
Kundu, M.~R. 1965, New York: Interscience Publication

\bibitem[{Landi {et~al.}(2002)Landi, Feldman, \& Dere}]{Landi2002}
Landi, E., Feldman, U., \& Dere, K.~P. 2002, The Astrophysical Journal, 574,
  495, \dodoi{10.1086/340837}

\bibitem[{Markwardt(2009)}]{markwardt2009non}
Markwardt, C.~B. 2009, arXiv preprint arXiv:0902.2850

\bibitem[{Mason {et~al.}(2016)Mason, Woods, Caspi, Chamberlin, Moore, Jones,
  Kohnert, Li, Palo, \& Solomon}]{Mason2016}
Mason, J.~P., Woods, T.~N., Caspi, A., {et~al.} 2016, Journal of Spacecraft and
  Rockets, 53, 328, \dodoi{10.2514/1.A33351}

\bibitem[{Mason {et~al.}(2020)Mason, Woods, Chamberlin, Jones, Kohnert, Schwab,
  Sewell, Caspi, Moore, Palo, Solomon, \& Warren}]{Mason2019}
Mason, J.~P., Woods, T.~N., Chamberlin, P.~C., {et~al.} 2020, Advances in Space
  Research, 66, 3, \dodoi{10.1016/j.asr.2019.02.011}

\bibitem[{Mazzotta {et~al.}(1998)Mazzotta, Mazzitelli, Colafrancesco, \&
  Vittorio}]{mazzotta1998ionization}
Mazzotta, P., Mazzitelli, G., Colafrancesco, S., \& Vittorio, N. 1998,
  Astronomy and Astrophysics Supplement Series, 133, 403

\bibitem[{Moore {et~al.}(2018)Moore, Caspi, Woods, Chamberlin, Dennis, Jones,
  Mason, Schwartz, \& Tolbert}]{moore2018}
Moore, C.~S., Caspi, A., Woods, T.~N., {et~al.} 2018, Solar Physics, 293,
  \dodoi{10.1007/s11207-018-1243-3}

\bibitem[{Richards {et~al.}(1994)Richards, Fennelly, \&
  Torr}]{richards1994euvac}
Richards, P., Fennelly, J., \& Torr, D. 1994, Journal of Geophysical Research:
  Space Physics, 99, 8981

\bibitem[{Richards {et~al.}(2006)Richards, Woods, \&
  Peterson}]{richards2006heuvac}
Richards, P.~G., Woods, T.~N., \& Peterson, W.~K. 2006, Advances in Space
  Research, 37, 315

\bibitem[{Sakurai(2017)}]{sakurai2017heating}
Sakurai, T. 2017, Proceedings of the Japan Academy, Series B, 93, 87

\bibitem[{Schmahl \& Kundu(1995)}]{schmahl1995microwave}
Schmahl, E., \& Kundu, M. 1995, Journal of Geophysical Research: Space Physics,
  100, 19851

\bibitem[{Schonfeld {et~al.}(2015)Schonfeld, White, Henney, Arge, \&
  McAteer}]{schonfeld2015coronal}
Schonfeld, S., White, S., Henney, C., Arge, C., \& McAteer, R. 2015, The
  Astrophysical Journal, 808, 29

\bibitem[{Schwab {et~al.}(2020)Schwab, Sewell, Woods, Caspi, Mason, \&
  Moore}]{schwab2020soft}
Schwab, B.~D., Sewell, R.~H., Woods, T.~N., {et~al.} 2020, The Astrophysical
  Journal, 904, 20

\bibitem[{Schwartz {et~al.}(2002)Schwartz, Csillaghy, Tolbert, Hurford,
  Mc~Tiernan, \& Zarro}]{schwartz2002rhessi}
Schwartz, R., Csillaghy, A., Tolbert, A., {et~al.} 2002, Solar Physics, 210,
  165

\bibitem[{Swarup {et~al.}(1963)Swarup, Kakinuma, Covington, Harvey, Mullaly, \&
  Rome}]{swarup1963high}
Swarup, G., Kakinuma, T., Covington, A., {et~al.} 1963, The Astrophysical
  Journal, 137, 1251

\bibitem[{Tapping(1987)}]{tapping1987recent}
Tapping, K. 1987, Journal of Geophysical Research: Atmospheres, 92, 829

\bibitem[{Tapping(2013)}]{tapping201310}
---. 2013, Space weather, 11, 394

\bibitem[{Tobiska {et~al.}(2008)Tobiska, Bouwer, \&
  Bowman}]{tobiska2008development}
Tobiska, W.~K., Bouwer, S.~D., \& Bowman, B.~R. 2008, Journal of Atmospheric
  and Solar-Terrestrial Physics, 70, 803

\bibitem[{Tolbert \& Schwartz(2020)}]{tolbert2020ospex}
Tolbert, K., \& Schwartz, R. 2020, Astrophysics Source Code Library, ascl

\bibitem[{Torr {et~al.}(1979)Torr, Torr, Ong, \&
  Hinteregger}]{torr1979ionization}
Torr, M.~R., Torr, D., Ong, R., \& Hinteregger, H.~E. 1979, Geophysical
  Research Letters, 6, 771

\bibitem[{Ware~Dewolfe {et~al.}(2010)Ware~Dewolfe, Wilson, Lindholm, Pankratz,
  Snow, \& Woods}]{ware2010solar}
Ware~Dewolfe, A., Wilson, A., Lindholm, D., {et~al.} 2010, in AGU Fall Meeting
  Abstracts, Vol. 2010, GC21B--0881

\bibitem[{Wild {et~al.}(1963)Wild, Smerd, \& Weiss}]{wild1963solar}
Wild, J., Smerd, S., \& Weiss, A. 1963, Annual Review of Astronomy and
  Astrophysics, 1, 291

\bibitem[{Woods \& Elliott(2022)}]{woods2022solar}
Woods, T.~N., \& Elliott, J. 2022, Solar Physics, 297, 1

\bibitem[{Yoshida \& Tsuneta(1996)}]{yoshida1996temperature}
Yoshida, T., \& Tsuneta, S. 1996, The Astrophysical Journal, 459, 342

\end{thebibliography}
\bibliographystyle{aasjournal}

\end{document}